# Stochastic network model of receptor cross-talk predicts anti-angiogenic effects

Amy L. Bauer*†, Trachette L. Jackson†, Yi Jiang* & Thimo Rohlf♦

*\* Theoretical Division, Los Alamos National Laboratory, Los Alamos, NM., USA.*

*† Department of Mathematics, University of Michigan, Ann Arbor, MI., USA.*

*♦ Santa Fe Institute, Santa Fe, NM., USA.*

**Cancer invasion and metastasis depend on tumor-induced angiogenesis, which is the formation of new blood vessels from existing vasculature in response to chemical signals from a tumor. The cellular processes (cell division, migration, and apoptosis) that occur during angiogenesis are tightly coordinated and regulated by signaling molecules. Thus, understanding how cells synthesize multiple biochemical signals initiated by key external stimuli can lead to the development of novel therapeutic strategies to combat cancer. An entire field of research is dedicated to the study of intracellular signaling cascades and biochemical reactions. In the face of large amounts of disjoint experimental data generated from multitudes of different laboratories using various assays, theoretical signal transduction models provide a framework to distill this vast amount of data. Such models offer an opportunity to formulate and test new hypotheses, and can be used to make experimentally verifiable predictions. This study is the first to propose a network model that highlights the cross-talk between the key receptors involved in angiogenesis, namely growth factor, integrin, and cadherin receptors. From available experimental data, we construct a stochastic Boolean network model of receptor cross-talk, and systematically analyze the dynamical stability of the network under continuous-time Boolean dynamics with a noisy production function. We find that the signal transduction network exhibits a robust and fast response to external signals, independent of the internal cell state. We derive an input-output table that maps external stimuli to cell phenotypes, which is extraordinarily stable against molecular noise with one important exception: an oscillatory feedback loop between the key signaling molecules RhoA and Rac1 is unstable under arbitrarily low noise, leading to erratic, dysfunctional cell motion. Finally, we show that the network exhibits an apoptotic response rate that increases with noise, suggesting that the probability of programmed cell death depends on cell health.**

During angiogenesis, the growth factor, integrin, and cadherin signaling pathways are highly connected and provide regulatory feedback to each other[1], a process referred to as receptor cross-talk. For example, in response to VEGF, endothelial cells upregulate the expression of integrin receptors[2]. Hutchings et al. 2003[3] found that integrins can additionally serve as receptors for immobilized $VEGF_{165}$ and $VEGF_{189}$ present in the extracellular matrix (ECM). Through the growth factor receptor (RTK), VEGF activates the MAPK signal transduction pathway stimulating proliferation and cell survival. Cell survival and proliferation, however, critically depend on adherence to the ECM, since even in the presence of stimulating concentrations of growth factor, loss of anchorage to the ECM results in cell cycle cessation and apoptosis[4,5]. Another example of receptor cross-talk in angiogenesis occurs through cadherin activation. Cadherins bind to actin, which is an important structural and signaling molecule for cytoskeleton reorganization. Therefore cadherin receptors not only facilitate cell-cell communication but also influence motility. Moreover, there is evidence that cadherins also induce signals that mitigate growth factor activation and repress cell proliferation[6]. This process is called contact inhibition. A cell interprets the coupled signals from growth factor, integrin, and cadherin receptors to determine cell phenotype and dynamically regulate angiogenic processes[7].

**Methods.**

**Continuous time Boolean networks: Deterministic dynamics.** Continuous time switching networks were introduced by Glass (1975)[8] as a differential equation model of gene expression dynamics. We now consider signaling networks; therefore we will talk more generally about "molecular species" instead of genes. Each molecular species, $i$, has a real-valued, normalized concentration level as a function of time, denoted $x_i(t) \in [0,1]$. Based on this concentration level, each molecular species is associated with a binary Boolean state $X_i(t)$ through a threshold switching mechanism:

$$X_i(t) = \begin{cases} 1, & \text{if } x_i(t) \geq 1/2 \\ 0, & \text{if } x_i(t) < 1/2. \end{cases}$$

As in Boolean networks, each species $i$ has $k_i$ regulators, $r_i^1,\ldots,r_i^{k_i}$ and a Boolean regulation function $f_i:[0,1]^{k_i} \to [0,1]$. The continuous time dynamics of species $i$ is given by the following stochastic differential equation:

$$\frac{dx_i(t)}{dt} = \left| f_i(X_{r_i^1}(t), \ldots, X_{r_i^{k_i}}(t)) - \delta(t) \right| - x_i(t). \tag{1}$$

Notice that Eqn. 1 has the form of a production-decay differential equation, hence $f_i(t)$ is referred to as production rate of molecular species $i$ at time $t$, and $f_i(t)$ plays the role of a production rate function. Since concentrations of signaling molecules in cells are usually relatively low, one can expect that a substantial amount of noise (randomness) occurs in molecular interactions. To account for this, we introduce the stochastic variable $\delta(t)$ in Eqn. 1, such that:

$$\delta(t) = \begin{cases} 1, & \text{with probability } p \\ 0, & \text{with probability } 1-p, \end{cases}$$

where $p \in [0,1/2]$. Notice that for $p=0$, the dynamics are completely deterministic, whereas $p=1/2$ corresponds to a complete randomization of the production function, i.e., making it a random switch. In between, $0 < p < 1/2$, the Boolean update executes with an error rate $p$, since when $\delta(t)=1$, the output determined by $f_i(t)$ is always inverted.

This study is the first to propose a signaling network model that highlights the cross-talk between growth factor, integrin, and cadherin receptors in angiogenesis. Fig. 1 graphically represents the simplified signal transduction network we implement for this study characterizing the key signaling pathways activated during angiogenesis. An arrow between nodes signifies activation and a hammerhead indicates an inhibitory effect. This signaling network is developed with the aim of synthesizing the empirical data available for endothelial cell signal transduction during critical angiogenic processes using the sparsest graph consistent with all experimental observations. We integrate data from signaling databases, including the KEGG Pathway Database[9], with results from experiments to determine the dependence relation for each signalling molecule in the network (Table 1). Since reaction rates for most of the kinetic interactions of interest are not available in the experimental literature, we employ a Boolean network model approach and show that a Boolean approach provides a reasonable description of the dynamics.

**Results.**

Using the model network shown in Fig. 1, we systematically test all possible combinations of input signals, with the internal network nodes initially set to randomly chosen concentration levels drawn from an equal distribution in [0,1]. From this starting state, dynamics are iterated for all network elements according to Eqn (1), while external signals are held constant. We find that the network always converges to a unique set of output states (phenotypes), independent from the initial internal state. Fig. 2 shows the resulting input-output table for the baseline network. The absence of only one of the two external stimuli (growth and motility signal) always induces apoptosis. When both growth and motility signals are present, cross-talk plays a crucial role in determining the actual phenotype of the cell. This can be clearly seen from the effect of the activation or deactivation of Rac1, a central player in mediating cross-talk in this network. Without contact inhibition, deactivation of Rac1 shunts cell growth; with Rac1 active, both cell growth and motility are present. For contact inhibited cells, this leads to the interesting finding that deactivtated Rac1 leads to quiescent cells, while activation of Rac1 produces a cell phenotype which is motile, but does not proliferate.

   Fig. 3 shows the response of the network to a sudden change in external stimulus, in this case the loss of the integrin signal. After the signal transduces through the network, an apoptotic response promptly follows. While the system is very sensitive to external signals, it is also very robust against fluctuations in concentrations. To test this, we vary the level of noise *p* in the production function between 0 and its maximum 0.5. We find that the network response (input-output map), as shown in Fig. 2, is reliably produced (100% ). This result holds even when we introduce an error rate of up to 35% in Boolean function execution. Above 35%, we find the transient appearance of "wrong" cell states, but even in this case the correct phenotype is produced most of the time. To further quantify the robustness of the signaling network dynamics, we also study the average phenotype error rate $e_p$ (averaged over all three outputs) as a function of *p*, for different time step sizes *dt*. Fig. 3b summarizes the findings. For all *dt < 1*, there is always a finite value $p_c$ below which $e_p$ vanishes. Decreasing *dt*, which is equivalent to increasing the concentrations of signaling molecules needed for signal amplification, shifts this transition towards *p = 0.5*. For intermediate *dt*, $e_p$ is between 0 and 0.5. For example, for *dt = 0.1*, which is equivalent to an average of only 10 molecules of each species, $p_c = 0.2$. This indicates a high robustness of signal transduction against molecular noise, even at very low molecular concentrations as is typically found in living cells.

Another interesting question is how cells respond to a sudden, transient increase in stress (a "shock"). We model this by a sudden, complete randomization of internal concentration levels, starting from a living cell that has both integrin and growth signal present, i.e., there is no external pressure to enter apoptosis. After the shock, applied for one time step, we find a transient increase of the probability to have an apoptotic response that, after a peak, goes down again. The average height of the peak, as well as the time needed for the decrease of apoptotic signal, increase with the noise rate *p*. If we interpret *p* as the amount of stress already present in the system, where low *p* indicates a very healthy state (high reliability of signal processing) and high *p ~ 0.5* a very unhealthy state, this observation means that cells become more and more likely to undergo programmed cell death after the shock the less healthy they are. Interestingly, to efficiently exploit this dependence biologically, the apoptotic switch must be slightly more sensitive compared to motility and growth signal switches in the sense that the cell's response should be triggered by a lower concentration of signaling molecules. For example, if the response is triggered at a threshold of 0.47 (dashed line in Fig. 4 lower panel), compared to 0.5 for growth and motility, cells with $p > 0.15$ would typically be induced to apoptosis, while cells with smaller $p$ would not respond. The selective value of this property is immediately evident: it implements a dynamical mechanism that makes sure that regeneration after a shock is started preferably from the most healthy cells, while unhealty cells are eliminated. Notice that a slight relative difference in equilibrium concentrations of signaling molecules involved in the respective cascades would have the same effect as a lowered threshold, which in the current model cannot be implemented since we normalize (maximal) concentrations. In principle, this is experimentally testable by comparing, e.g., the expression levels of the genes coding for signal molecules involved in apoptosis or growth/motility respectively.

Since Rac1 and RhoA are key players in cross-talk between integrin and growth signals, a large bulk of experimental work is concerned with the interplay between these molecules [13,39,48,62]. Figure 1b shows three different interaction schemes that are discussed in the literature. Interestingly, we find in simulations that both the baseline circuit (bl) without feedback, and the negative feedback circuit (fb 1) lead to stationary dynamics which is highly robust against noise, whereas a mixed feedback scheme (fb 2), where Rac1 activates RhoA and RhoA inhibits Rac1, is extremely sensitive to noise (while, for zero noise, it behaves as a highly regular oscillator). At the output (phenotype) level, given growth factor and integrin signals with low Rac1 activity, this leads to an erratic, dysfunctional on-off pattern of cell motility (Fig. 4, upper panel) even for very low levels of noise, indicating a complete breakdown of motility regulation. Since robustness against fluctuations of molecular concentrations is a key requirement for living cells, we predict that, with a high probability, either the feedback circuit fb 2 is not realized in healthy cells, or additional regulatory interactions must be present, which are not captured in the current model, to suppress these strongly nonlinear effects.

**Discussion.**

A major challenge facing the experimental research and the modelling communities is to integrate the vast amount of available information in a way that improves our understanding of the principal underpinnings driving angiogenic processes and that will advance efforts aimed at the development of new therapies for treating cancer and other angiogenesis-dependent diseases. Driven by the scarcity of quantitative kinetic data for the biochemical reactions of interest, we develop a stochastic Boolean network model to describe the signal transduction pathways critical to cellular regulation and function during angiogenesis. This model is the first to couple growth factor, integrin, and cadherin signaling cascades in receptor cross-talk during angiogenesis in a manner consistent with experimental observations. Using this model, we identify relationships between receptor activation combinations and cellular function, and show that receptor cross-talk is crucial to cell phenotype determination. Specifically, we show which combinations of Rac1 (de)activation and cadherin regulated contact inhibition control cell quiescence, growth, and motility. In addition, we study the controversial relationship between RhoA and Rac1 and predict that, due to the signalling instability created when Rac1 activates RhoA, but RhoA inhibits Rac1, that this is an unlikely feedback mechanism in healthy cells or that additional regulatory interactions must be present. We also find evidence that the apoptotic switch is dependent on cell health and that the sensitivity of the apoptotic switch increases with the amount of stress (noise) already present in the system, suggesting a dynamical mechanism for cell survival selection whereby cells that are already under stress require a lower concentration of signaling molecules to trigger apoptosis. These results translate and synthesize a large body of compartmentalized research on molecular signaling pathways to increase our understanding of the molecular mechanisms associated with the regulation of cell growth, motility, quiescence, and apoptosis, and suggest specific molecules and relationships that can be targeted for therapeutic gain.

Correspondence and requests for materials should be addressed to ALB (e-mail: albauer@lanl.gov).

Figure 1. Receptor Cross-talk During Angiogenesis. Simplified signal transduction network linking external stimuli to a cell's internal decision making machinery. This network highlights the relationship between VEGF, integrin, and cadherin receptors, allowing for cross-talk between the three to ultimately decide the cell's fate. An arrow between nodes signifies activation and a hammerhead indicates an inhibitory effect.

Figure 2. Network Predicts Cell Phenotype Depends on External Signals. Table summarizing cell phenotype predictions by the Boolean network for various input configurations, where, for example, [10] denotes a VEGF signal only and output (100) indicates proliferation. This Boolean network model exhibits five critical and distinct cell phenotypes: apoptotic, proliferating, migrating, quiescent, and both proliferating and migrating.

Figure 3a. Robust and Rapid Network Response. Left: Screenshot of system dynamics, with external inputs left, outputs right (orange), time runs from top to bottom. White arrow indicates that ITG signal is turned off; orange arrow indicates the output response (apoptosis). Right: Output response (real valued concentration level) as a function of time. After the signal has passed through the network, there is a quick apoptotic response. Results are obtained using a noise level *p = 0.1*.

Figure 3b. Network is Robust to Noise in Internal Signals. Average phenotype error rate (i.e. fraction of "wrong" output states), averaged over 10,000 iterations of the discretized set of differential equations (Eqn. 1), for three different time step sizes (*dt = 0.1, dt = 0.01, dt = 0.001*, from left to right). For all *dt < 1*, there is a sharp transition at a finite value $p_c$ of the error rate $p_e$, below which phenotype errors vanish. Note that decreasing *dt* is equivalent to increasing the average concentration of signalling molecules.

Figure 4a. Negative Rac1/RhoA Feedback Loop Predicted. Motility response given RTK and ITG signals with low Rac1 activity at a noise rate *p = 0.1*. Both the baseline network and the network where RhoA and Rac1 inhibit each other (feedback scheme 1) lead to proper regulation and inhibit motility (signal well below 0.5). If Rac1 activates RhoA (feedback scheme 2), even low levels of noise lead to an erratic output signal with strong fluctuations around 0.5, and hence an erratic on-off pattern of motility. Since this behavior is not biologically functional, we postulate that, with high probability, feedback scheme 2 is ***not*** realized in this biological system.

Figure 4b. Apoptosis Response Sensitive to Noise or Apoptosis Regulation for Angiogenic Effects. Transient apoptotic signal for RTK and ITG input with Rac1 activated. Curves averaged over 1,000 different initial conditions of the internal network. The apoptosis signal increases as the noise level, p, increases, suggesting that the signal may have a biological function: when noise rates become to high, e.g., due to cell stress, apoptosis can be triggered before cells proliferate or move. This requires that the apoptotic switch is slightly more sensitive than the other switches: e.g., with a threshold of 0.47 (dashed line), any noise level *p > 0.15* would trigger apoptosis.

**Table of Boolean network interactions.** Table summarizing the Boolean dependence relation for each node determined based on current scientific literature. References are given. Node numbers correspond to nodes in Figure 1a and $s_i$ denotes the state (0 or 1) of node $i = \{0, ..., 22\}$ at time $t-1$. A zero indicates the signal is off and a one indicates the signal is on. For instance, Ras (node 3) is activated if Grb-2 (node 2) is, whereas GSK-3β (node 11) is inhibited when Akt (node 9) is activated. Most of this information is specific to endothelial cells, but data from other cells lines are included where information on the endothelial cell line is lacking. During simulations, internal nodes are initialized to $s = 1$ with probability $p = 0.5$, otherwise the node assumes $s = 0$.

| Node # | Dependence Function | Reference |
|---|---|---|
| -1 | external signal (VE-cadherin contact inhibition) | [6] |
| 0 | external signal (VEGF binding) | [10] |
| 1 | external signal (integrin binding) | [11] |
| 2 | s0 | [10] |
| 3 | s2 | [6] |
| 4 | s3 | [12] |
| 5 | s4 OR s10 | [13] |
| 6 | s5 | [12] |
| 7 | s3 AND s14 | [11,12,14] |
| 8 | s7 AND NOT s16 | [15,16] |
| 9 | s8 | [14] |
| 10 | s7 AND NOT s17 (inactive) (OR if activated) | [15,17,18] |
| 11 | NOT s9 | [14] |
| 12 | NOT s9 | [12] |
| 13 | NOT s11 | [14] |
| 14 | s1 | [19] |
| 15 | NOT s1 | [20] |
| 16 | s15 | [21] |
| 17 | s14 | [9] |
| 18 | s17 AND NOT s6 | [20] |
| 19 | s18 OR s10 | [22,23] |
| 20 | s6 OR s13 | [24] |
| 21 | s12 AND NOT s13 | [14] |
| 22 | s19 | [25] |

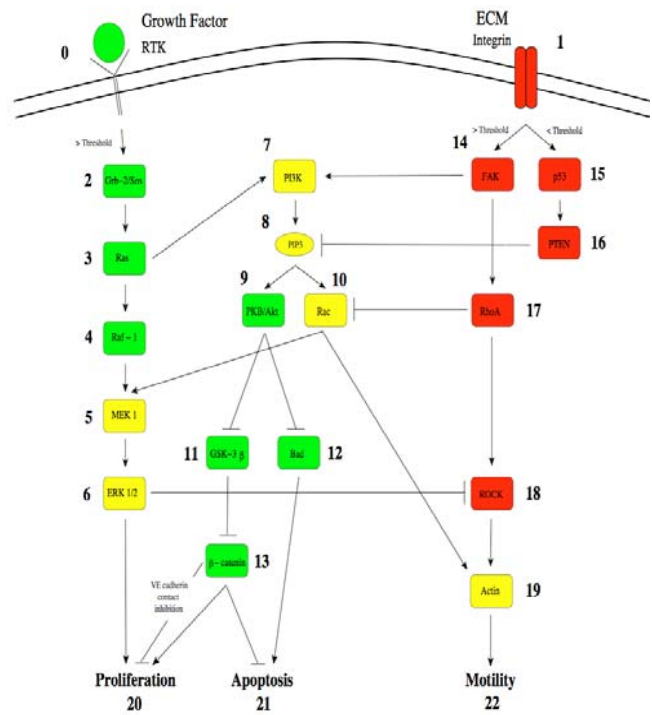

Figure 1a

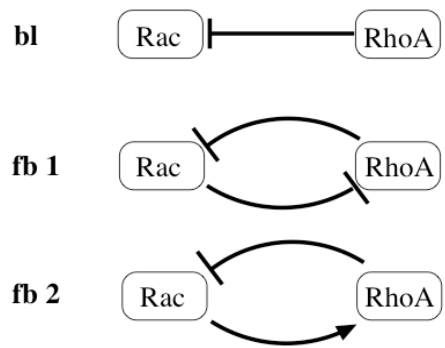

Figure 1b

|  | [RTK, ITG] Input State | | | |
|---|---|---|---|---|
|  | [0 0] | [1 0] | [0 1] | [1 1] |
| Off | 0 1 0 | 0 1 0 | 0 1 1 | 1 0 0 |
| On | 0 1 1 | 0 1 1 | 0 1 1 | 1 0 1 |
| Off | 0 1 0 | 0 1 0 | 0 1 1 | 0 0 0 |
| On | 0 1 1 | 0 1 1 | 0 1 1 | 0 0 1 |

Rac Activation State (rows 1–2 and 3–4); Contact Inhibition (off for top two rows, on for bottom two rows)

Figure 2

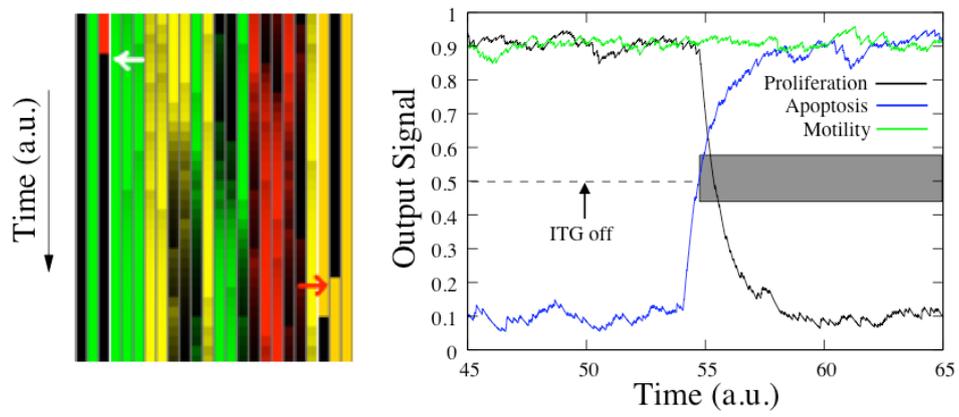

Figure 3a

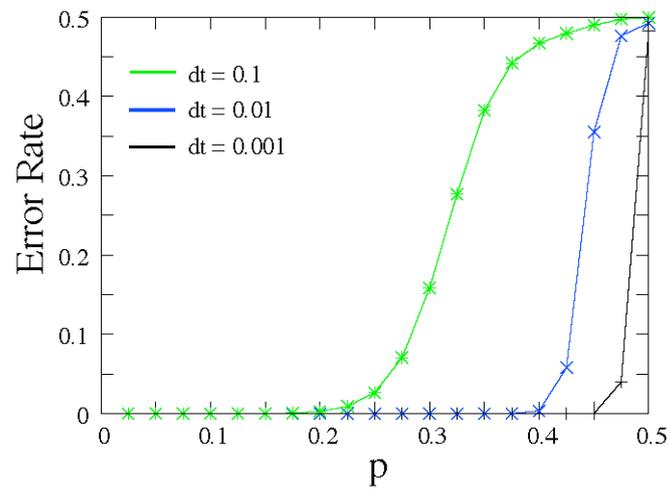

Figure 3b

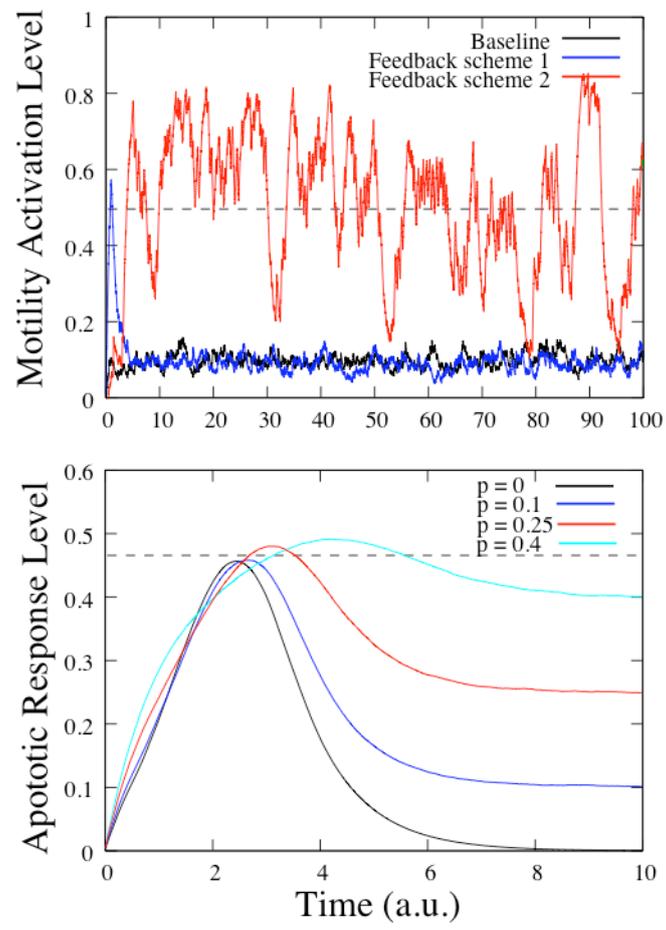

Figure 4